\documentclass[12pt]{iopart}

\usepackage{graphics}

\renewcommand{\d}{\mathrm{d}}
\newcommand{\dfrac}{\frac}
\newcommand{\tx}{\xi}

\begin{document}

\title[Circular geodesics in the equatorial plane of an extreme Kerr-Newman black hole]
{A note on circular geodesics in the equatorial plane of an extreme Kerr-Newman black hole}

\author{Sebastian Ulbricht and Reinhard Meinel}

\address{Theoretisch-Physikalisches Institut,
University of Jena,\\
 Max-Wien-Platz 1, 07743 Jena, Germany}

\ead{sebastian.ulbricht@uni-jena.de, meinel@tpi.uni-jena.de}
\begin{abstract}
We examine the behaviour of circular geodesics describing orbits of neutral test particles around an extreme 
Kerr-Newman black hole.
It is well known that the radial Boyer-Lindquist coordinates of the prograde photon orbit $r=r_{\rm ph}$, marginally 
bound orbit $r=r_{\rm mb}$ and innermost stable orbit $r=r_{\rm ms}$ of the extreme Kerr black hole all coincide with 
the event horizon's value $r=r_+$. We find that for the extreme Kerr-Newman black hole with 
mass $M$, angular momentum $J$ and electric charge $Q=\pm\sqrt{M^2-J^2/M^2}$ ($|J|\le M^2$) the coordinate equalities
$r_{\rm ph}=r_+$, $r_{\rm mb}=r_+$ and $r_{\rm ms}=r_+$ hold if and 
only if $|J|$ is greater than or equal to $M^2/2$, $M^2/\sqrt{3}$ and $M^2/\sqrt{2}$, respectively.
\end{abstract}

\pacs{04.20.-q, 04.70.Bw}

\vspace{2pc}

\section{Introduction}
A classic publication by Bardeen \emph{et al.} \cite{B72} gives a detailed analysis of circular geodesics in the 
equatorial plane of a Kerr black hole. The equations leading to the marginally stable and marginally bound orbits 
as well as the photon orbit for the Kerr-Newman spacetime were given by Dadhich and Kale \cite{DK} 
in terms of the Boyer-Lindquist coordinate $r$. The equation for the photon 
orbit is of 4th order in $\sqrt{Mr-Q^2}$ and can thus be solved analytically.
For the extreme Kerr-Newman solution, a very simple expression for the photon orbit follows, which can be 
found in the work by Balek \emph{et al.} \cite{BBS}. 
We also discuss the solutions for the marginally bound and marginally stable orbits. It turns 
out that in the prograde case, all these radii coincide with the horizon's value $r_+$ for all angular momenta of 
the black hole greater than a special value. Such an agreement in coordinate values is well known from the extreme 
Kerr metric. We note that it does not mean that the locations of these orbits and the horizon are the same on slices of
constant Boyer-Lindquist time, but they are the same on horizon-crossing slices as was shown by Jacobson \cite{J}.

\section{Kerr-Newman metric in the equatorial plane}
We start with the Kerr-Newman metric in Boyer-Lindquist  coordinates $(r,\theta,\phi,t)$:
\begin{equation*}
\hspace{-1cm}\d s^2=\dfrac{\Sigma}{\Delta}\d r^2 + \Sigma \d \theta^2 +\left(r^2+(J/M)^2+
\dfrac{(2Mr-Q^2)(J/M)^2\sin^2\theta}{\Sigma}\right)\sin^2\theta \,\d\phi^2
\end{equation*}
\begin{equation}
\phantom{\d s^2=}-\dfrac{2\,(2Mr-Q^2)(J/M)\sin^2\theta}{\Sigma}\,\d\phi\,\d t - 
\left(1-\dfrac{2Mr-Q^2}{\Sigma}\right)\,\d t^2\,,
\end{equation}
\begin{equation}
\textnormal{with} \quad \Delta=r^2 -2Mr+(J/M)^2+Q^2\,,\quad \Sigma= r^2+(J/M)^2\cos^2\theta
\end{equation}
and where we have set the gravitational constant and the speed of light equal to one.
From here on in we study the equatorial plane $\theta=\pi/2$. We use the coordinates per mass as a characteristic 
length ($\d s'=\d s/M$,  $x=r/M$, $t'=t/M$) and rescale the parameters ($q=Q/M$, $a=J/M^2$).  
All expressions are dimensionless during the calculation.  
The corresponding parameter space is $\{(a,q)\,|\,a^2+q^2\leq 1\,\textnormal{and}\,\, a,q\geq 0\}$. 
(Without loss of generality we assume non-negative $Q$ and $J$.) The metric then becomes
\begin{equation}
\d s'^2=g_{11}\,\d x^2 + g_{33}\, \d \phi^2 + 2 g_{34}\, \d \phi\d t' + g_{44}\, \d t'^2\,,
\end{equation}
with the metric coefficients
\begin{equation}
g_{11}=\frac{x^2}{x^2-2x+a^2+q^2}\,,
\end{equation}
\begin{equation}
g_{33}=x^2+a^2+\frac{(2x-q^2)a^2}{x^2}\,,
\end{equation}
\begin{equation}
g_{34}=-(2x-q^2)\frac{a}{x^2}\,,
\end{equation}
\begin{equation}
g_{44}=-\frac{x^2-2x+q^2}{x^2}\,.
\end{equation}
We note that the event horizon of the black hole is characterized by 
\begin{equation}
 x=x_+=1+\sqrt{1-a^2-q^2}\,.
\end{equation}

\section{Constants of motion and angular velocity}
Since the metric does not depend on $t'$ and $\phi$, we can find two constants $A$ and $B$, 
associated with the specific energy and the specific angular momentum of a particle 
(also discussed in detail in \cite{B72, DK}): 
\begin{equation}
A=-g_{44}\dot t' - g_{34} \dot \phi\,,\quad B=g_{33}\dot\phi+g_{34}\dot t'  \label{eqn:AandB}\,,
\end{equation}
where the dot denotes a differentiation with respect to proper time. For circular orbits, these constants can be 
determined by analyzing the zeros of an effective potential $V(x)$ and its derivative $V'(x)$.
During the calculation it is helpful to introduce the new coordinate $\tx=x-q^2$ instead of $x$, to handle 
polynomials of $\tx^{1/2}$ instead of combinations of $x$ and $\sqrt{x-q^2}$. The results are:
\begin{equation}
A^{\pm}_{\mathrm{KN}}(\tx,a,q)=\phantom{\pm}\frac{q^4+q^2(2\tx -1)+\tx^2-2\tx\pm a\tx^{1/2}}{(\tx+q^2)[q^4+q^2(2\tx -1)+
\tx^2-3\tx\pm 2a\tx^{1/2}]^{1/2}} \label{eqn:AKN,pm}\,,
\end{equation}
\begin{equation}
B^{\pm}_{\mathrm{KN}}(\tx,a,q)=\pm\frac{a^2\tx^{1/2}+\tx^{1/2}(\tx+q^2)^2\mp a(2\tx + q^2)}{(\tx+q^2)[q^4+q^2(2\tx -1)+
\tx^2-3\tx\pm 2a\tx^{1/2}]^{1/2}} \label{eqn:BKN,pm}\,.
\end{equation}
The upper sign stands for prograde and the lower one for retrograde circular orbits around the black hole.\smallskip\\
The angular velocity, given by $\Omega=\d \phi/ \d t'$, can be found by combining equations (\ref{eqn:AandB}):
\begin{equation}
\Omega = - \frac{g_{44}B+g_{34}A}{g_{33}A+g_{34}B}\,.
\end{equation}
With the equations (\ref{eqn:AKN,pm}), (\ref{eqn:BKN,pm}) a very short expression for $\Omega$ follows:
\begin{equation}
\Omega^\pm_{\mathrm{KN}}(\tx,a,q)=\pm\frac{\tx^{1/2}}{(q^2+\tx)^2\pm a \tx^{1/2}} \label{eqn:OmegaKN,pm}\,.
\end{equation}
\section{Behaviour of the photon orbit in the extreme case}
\begin{figure}[ht]
\begin{center}\includegraphics{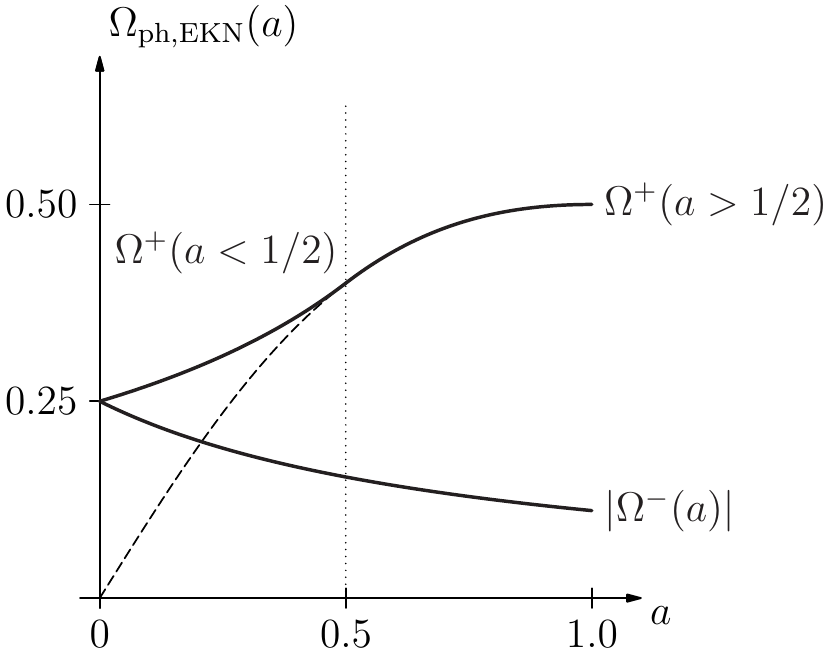}\end{center}
\caption{The angular velocity as a function of $a \in\, [0,1]$ at the photon orbit. \\
The dashed line starting at zero is the angular velocity of the horizon (at $x=1$).}
\label{fig1}
\end{figure}

The photon orbit is the only light-like circular geodesic around the black hole, hence the name. 
In addition it is a limiting case for the innermost time-like geodesic. A particle with non-zero rest mass 
would need an infinite amount of energy to be on this orbit. In this case $A^{\pm}_{\mathrm{KN}}$ must diverge and we 
find the photon orbit $x^{\pm}_{\mathrm{ph,KN}}(a,q)$ as the outer solution of
\begin{equation}
q^4+q^2(2\tx -1)+\tx^2-3\tx\pm 2a\tx^{1/2}\stackrel{!}{=}0 \label{eqn:xphKN,pm}\,.
\end{equation}
This polynomial is of 4th order in $\tx^{1/2}$ and has an  analytical solution for the whole parameter 
space $\{(a,q)|\,\,a^2+q^2\leq 1\,\textnormal{and}\,\, a,q\geq 0\}$. 
The most interesting part of the parameter space is the extremal case $a^2+q^2=1$, leading to the condition
\begin{equation}
(\tx^{1/2}\mp a)^2(\tx^{1/2}-1\pm a)(\tx^{1/2}+1\pm a)\stackrel{!}{=}0 \label{eqn:xphEKN,pm}\,,
\end{equation}  
where we use $a$ as the remaining free parameter.
We see that $\tx=a^2$ is a solution for prograde orbits. That is equivalent to the horizon's coordinate  $x=1$ of 
the extreme Kerr-Newman spacetime. The other solutions are $\tx^{1/2}=1\mp a$, leading to:
\begin{equation}
x^{+}_{\mathrm{ph,EKN}}(a)=\left\{\begin{array}{ll}
2(1-a) &\quad,\,a<1/2\\
1 &\quad,\,a>1/2\,\,,
\end{array}\right.
\end{equation}
\begin{equation}
x^{-}_{\mathrm{ph,EKN}}(a)=\hspace{0.45cm}2(1+a)\,.
\end{equation}
These formulae were already derived in \cite{BBS}, see equations (2.15) of that paper.
For $a=0$ we find the photon orbit $x=2$ of the extreme Reissner-Nordstr{\o}m solution. 
For the prograde orbit, there is a change in the orbit's behaviour at $a=1/2$.
The angular velocity of the photon orbit is:
\begin{equation}
\Omega^{+}_{\mathrm{ph,EKN}}(a)=\left\{\begin{array}{ll}\dfrac{1}{4-3a} &\quad,\,a<1/2\smallskip\\
\dfrac{a}{1+a^2} &\quad,\,a>1/2\,\,,
\end{array}\right.
\end{equation}
\begin{equation}
\Omega^{-}_{\mathrm{ph,EKN}}(a)=-\dfrac{1}{4+5a}\,.
\end{equation}
The angular velocity of the prograde photon orbit connects to the horizon's angular velocity in a continuously
differentiable way (with respect to the parameter $a$), see Figure \ref{fig1}. 

\section{Marginally stable and marginally bound orbit in the extreme case}
\begin{figure}[ht]
\begin{center}
\includegraphics{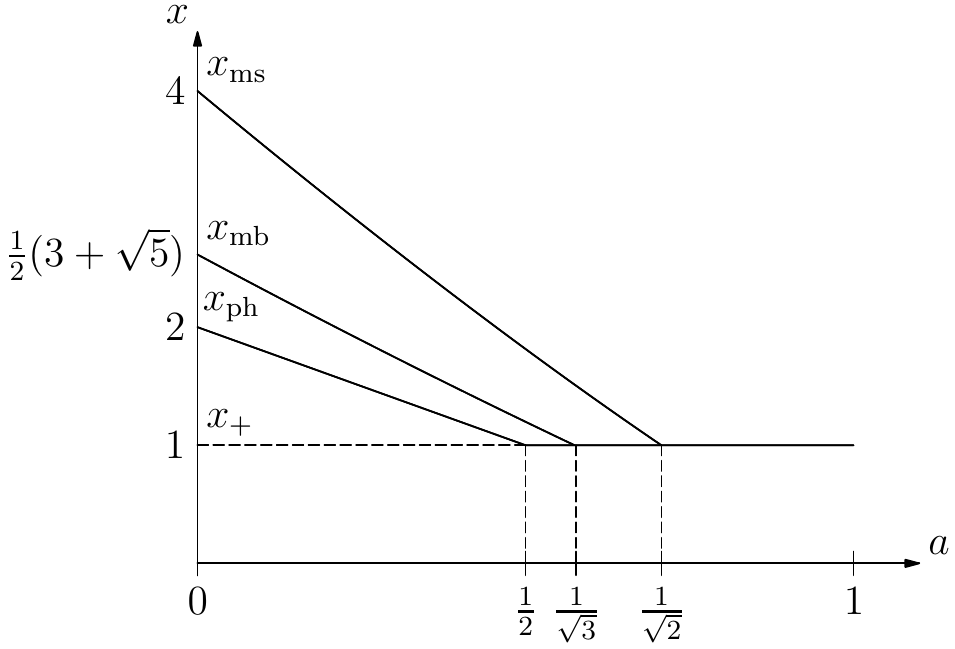}
\end{center}
\caption{The prograde photon orbit, marginally bound orbit and marginally stable orbit in 
the extreme Kerr-Newman spacetime as a function of the angular momentum parameter $a=J/M^2$. 
They all start with the values for the extreme Reissner-Nordstr{\o}m solution and coincide 
with the horizon's value $x=1$ from a special value of $a$ on.}
\label{fig2}
\end{figure}

If the second derivative of the effective potential with respect to $x$ changes its sign, the associated marginally 
stable orbit or ISCO (innermost stable circular orbit) $x^{\pm}_{\mathrm{ms,KN}}(a,q)$ defines the border between 
stable and unstable circular orbits. This leads to a 6th order polynomial in $\tx^{1/2}$:
\begin{equation}
q^2 (a^2 - q^2 + q^4) - 3 (a^2 + q^2 - q^4) \tx \pm 8 a \tx^{3/2} + 
 3 (-2 + q^2) \tx^2 + \tx^3\stackrel{!}{=}0\,. \label{eqn:xmsKN,pm}
\end{equation}
In the extremal case  this leads to a 3rd order polynomial in $\tx^{1/2}\pm a$:
\begin{equation}
(\tx^{1/2}\mp a)^3 [(\tx^{1/2}\pm a)^3-3(\tx^{1/2}\pm a)\pm 2a]\stackrel{!}{=}0 \label{eqn:xmsEKN,pm}\,,
\end{equation}
leading to the marginally stable orbit:
\begin{equation}
x^{+}_{\mathrm{ms,EKN}}(a)=\left\{\begin{array}{ll}1-a^2+\left[2\cos\left[1/3\,\arccos(-a)\right]-a\right]^2 &,\, 
a<1/\sqrt{2}\\
1&,\, a>1/\sqrt{2}\,\,,\end{array}\right.
\end{equation}
\begin{equation}
x^{-}_{\mathrm{ms,EKN}}(a)=1-a^2+\left[2\cos\left[1/3\,\arccos(a)\right]+a\right]^2\,.
\end{equation}
From $a=1/\sqrt{2}$ on, the ISCO and horizon values coincide.  \smallskip\\
A similar discussion of the marginally bound orbit $x^{\pm}_{\mathrm{mb,EKN}}(a)$ 
(with the condition $A^\pm_{\mathrm{EKN}}\stackrel{!}{=}1$) gives rise to a 6th order polynomial\footnote{This is 
equivalent to the corresponding equation (17) of \cite{DK}. Note that the first plus sign in the second line of 
that expression has to be a minus sign.} in $\tx^{1/2}$: 
\begin{equation}
-q^4 + q^6 \pm 2 a q^2 \tx^{1/2} - (a^2 + 4 q^2  - 3 q^4) \tx \pm
 4 a \tx^{3/2} -( 4 - 3 q^2 )\tx^2 + \tx^3\stackrel{!}{=}0
\end{equation}
reducing to a 4th order polynomial in the extreme case:
\begin{equation}
(\tx^{1/2}\mp a)^2 [\tx^{2}\pm 2a\tx^{3/2}-\tx\pm 2a(1-a^2)\tx^{1/2}-(1-a^2)^2]\stackrel{!}{=}0 \label{eqn:xmbEKN,pm}\,,
\end{equation}
and a value of $a=1/\sqrt{3}$ for the starting point of where the orbit and horizon values coincide. 
We note that the angular velocities 
corresponding to $x_{\rm mb}$ and $x_{\rm ms}$ are continuous but not differentiable with respect to $a$ at 
$a=1/\sqrt{3}$ and $1/\sqrt{2}$, respectively. The results for the characteristic radii 
are summarized in Figure \ref{fig2}.

\ack{We thank David Petroff for carefully reading a previous version of the manuscript and 
making valuable suggestions for amendments.}


\section*{References}
\begin{thebibliography}{10}
\bibitem{B72} Bardeen J M, Press W H and Teukolsky S A 1972 {\it Astrophys. J.} {\bf 178} 347-369
\bibitem{DK} Dadhich N and Kale P P 1976 {\it J. Math. Phys.} {\bf 18}
1727-1728 
\bibitem{BBS} Balek V, Bi\v{c}\'{a}k J and Stuchl\'{i}k Z 1989 {\it Bull. Astronom. Inst. Czechoslovakia} 
{\bf 40} 133-165
\bibitem{J} Jacobson T 2011 {\it Class. Quantum Grav.} {\bf 28} 187001
\end{thebibliography}
\end{document}